\documentclass[prl,twocolumn,showpacs,superscriptaddress,amsmath,amssymb]{revtex4}
\usepackage{graphicx}
\usepackage{dcolumn}
\usepackage{bm}
\usepackage{multirow}

\newcommand{\Cl}{$\kappa$-(ET)$_2$Cu[N(CN)$_2$]Cl}
\newcommand{\kClstar}{$\kappa$-Cl$^\ast$}
\newcommand{\kCl}{$\kappa$-Cl}

\begin{document}

\preprint{APS/123-QED}

\title{Different superconducting percolation regimes in the layered organic Mott system $\kappa$-(BEDT-TTF)$_2$Cu[N(CN)$_2$]Cl -- evidence from fluctuation spectroscopy}
\author{Jens M\"uller}
 \email{mueller@cpfs.mpg.de}
% \altaffiliation[Present address: ]{Physikalisches Institut, Goethe-University Frankfurt, 60438 Frankfurt(M), Germany}
\author{Jens Brandenburg}
\affiliation{Max Planck Institute for Chemical Physics of Solids, N\"othnitzer-Str.\ 40, 01187 Dresden, Germany}
\author{John A.\ Schlueter}
\affiliation{Argonne National Laboratory, Materials Science Division, Argonne, IL 60439, USA}

\date{\today}

\begin{abstract}
Fluctuation spectroscopy is used to investigate the organic bandwidth-controlled Mott system $\kappa$-(BEDT-TTF)$_2$Cu[N(CN)$_2$]Cl. We find evidence for percolative-type superconductivity in the spatially inhomogeneous coexistence region of antiferromagnetic insulating and superconducting states. % which develops under pressure at low temperatures.
When the superconducting transition is driven by a magnetic field, percolation seems to be dominated by unstable superconducting clusters upon approaching $T_c (B)$ from above, before a "classical" type of percolation is resumed at low fields, dominated by the fractional change of superconducting clusters. The $1/f$ noise is resolved into Lorentzian spectra in the crossover region where the action of an individual mesoscopic fluctuator is enhanced.  
\end{abstract}


\maketitle
\begin{figure}[]
\includegraphics[width=.475\textwidth]{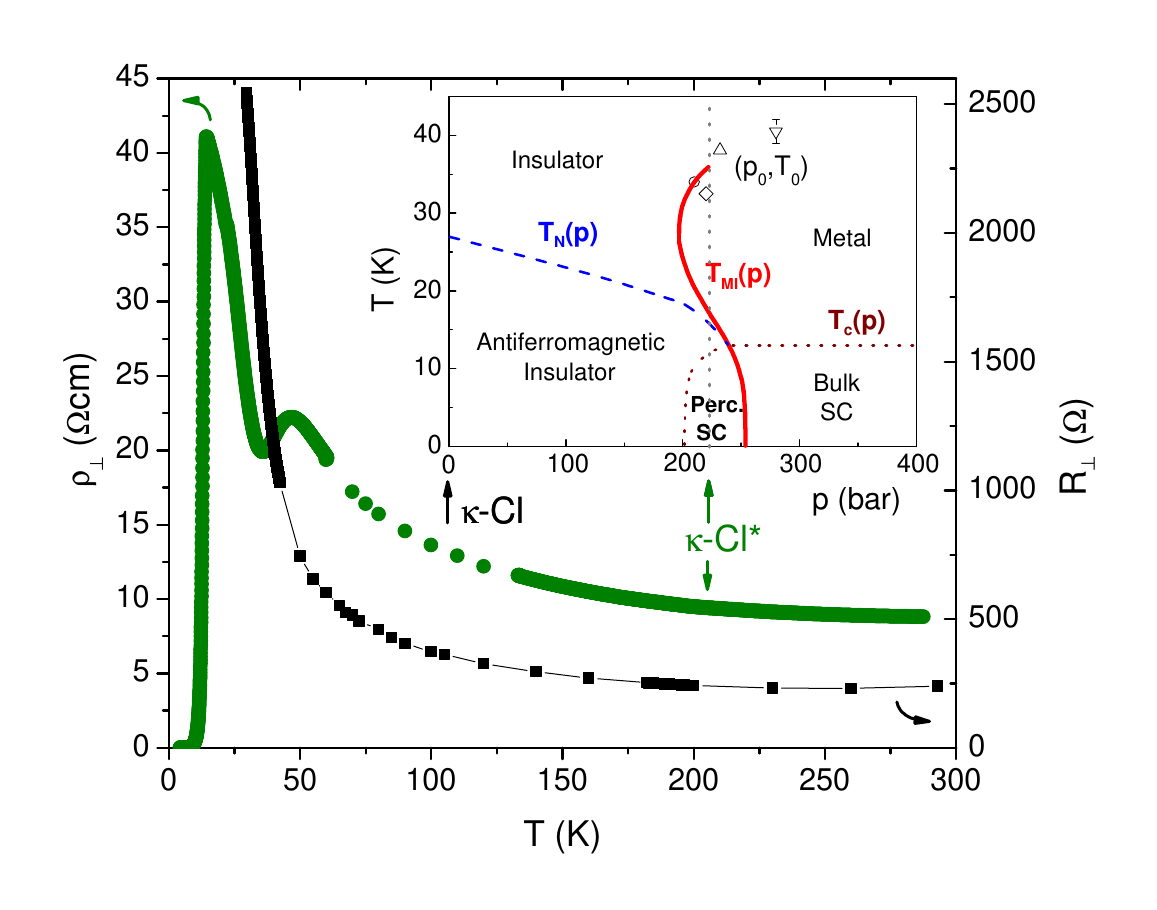}
\caption[]{(Color online.) Resistivity (resistance) of two samples of $\kappa$-(ET)$_2$Cu[N(CN)$_2$]Cl. Inset shows the experimental phase diagram after \cite{Lefebvre2000,Limelette2003,Fournier2003,Kagawa2004,deSouza2007}. $T_N$ and $T_c$ denote the N$\acute{{\rm e}}$el and superconducting transition temperatures, respectively. Arrows indicate the positions of the samples measured at ambient conditions (\kCl) and at a finite pressure (\kClstar).}\label{figure1}
\end{figure}

Materials in which magnetic, correlated metallic and superconducting phases compete for stability often show an intrinsic tendency to electronic phase separation, an effect which is intensively discussed for the high-$T_c$ cuprates and the manganites, see \cite{Kivelson2003,Dagotto2003} for reviews. For these correlated electron systems based on oxide materials, {\em nano-scale} inhomogeneities are reported including stripes and variations in the superconducting gap in high-$T_c$ materials \cite{Tranquada1995,KMLang2002} or the phase separation in doped manganese oxides \cite{Faeth1999}. 
Recently, also for quasi-two-dimensional organic conductors $\kappa$-(BEDT-TTF)$_2$X, where BEDT-TTF (or simply ET) is bis(ethylenedithio)tetrathiafulvalene and X stands for polymeric anions \cite{Toyota2007}, an inhomogeneous coexistence of antiferromagnetic insulating and superconducting phases is observed when the material is located in the vicinity of the first-order metal-to-insulator transition (MIT), which can be tuned either by altering the chemical composition of the material or by applying pressure \cite{Ito1996,Lefebvre2000,Miyagawa2002,Kagawa2004,Sasaki20045}. 
Pressure studies of the compound with X = Cu[N(CN)$_2$]Cl (\kCl\ hereafter) revealed a first-order MIT line $T_{MI}(p)$ \cite{Ito1996,Lefebvre2000,Limelette2003,Fournier2003,Kagawa2004}, indicative of a bandwidth-controlled Mott transition. The region of inhomogeneous phase coexistence in \kCl\ has been studied by simultaneous measurements of $^{1}$H-NMR and ac-susceptibility \cite{Lefebvre2000}, $^{13}$C-NMR \cite{Miyagawa2002}, and resistivity \cite{Ito1996,Kagawa2004}. Despite strong evidence for spatially and statically coexisting phases, the question was raised whether the coexistence was macroscopic or of mesoscopic type \cite{Lefebvre2000}. The details of spatial distribution, domain size, and stability of the inhomogeneous state are still under debate \cite{Sasaki20045}.
From resistance measurements, a percolative superconducting phase was suggested where tiny superconducting domains are induced progressively in the insulating host phase at small enough pressures \cite{Kagawa2004}, see phase diagram in Fig.\,\ref{figure1}.
In contrast, in partially deuterated samples of the material with X= Cu[N(CN)$_2$]Br ($\kappa$-Br) metallic and insulating domains of macroscopic size ($50 - 100\,\mu$m) were found in real-space imaging employing scanning microregion-infrared spectroscopy \cite{Sasaki20045} below the critical temperature $T_0 = 30 - 40$\,K of the first-order Mott transition.\\ In this Letter we report fluctuation spectroscopy measurements on bulk single crystals of the title organic charge-transfer salt aiming to investigate the percolative nature of the supercondicting transition in pressurized \kCl. We find evidence for different percolation regimes when the superconducting transition is driven by a magnetic field, and low-frequency switching of an individual {\em mesoscopic} fluctuating entity.\\
Single crystals of \Cl\ were grown by electrochemical crystallization as described elsewhere \cite{Williams1990}. Samples were of plate- %[$(0.6 \times 0.8 \times 0.2)\,{\rm mm}^3$, 
and rod-like morphology with the smallest and longest dimension, respectively, perpendicular to the conducting planes.
%smallest dimension perpendicular to the conducting planes] and rod-like morphology [$(0.36 \times 0.36 \times 1.2)\,{\rm mm}^3$, longest dimension perpendicular to the planes]. 
In order to produce a small pressure, samples (denoted as $\kappa$-Cl$^\ast$ hereafter) have been embedded in a solvent-free epoxy, the slightly larger coefficient of thermal expansion of which results in a finite stress acting on the sample during cooldown. Although the exact pressure is not known {\it a priori}, the actual conditions have been found to be reproducible for different cooldowns. An effective pressure of $\sim 220 - 250$\,bar on our samples has been estimated from the comparison to resistivity curves obtained from He-gas pressure experiments reported in the literature \cite{Kagawa2004}. These values are in good agreement with the effect on a superconducting reference compound for which the (known) pressure dependence of $T_c$ has been used as pressure "gauge".  
Low-frequency resistance fluctuations have been measured by a standard bridge-circuit ac-technique \cite{Scofield1987} in a five- or six-terminal (\kClstar) or four-terminal (\kCl) setup. The output signal of a lock-in amplifier (SR830) operating at a driving frequency $f_0$ of typically 517\,Hz was processed by a spectrum analyzer (HP35660A). Care was taken to exclude that spurious noise sources contributed to the results.  

Fig.\,\ref{figure1} compares the temperature dependence of the resistivity of \kClstar\ with that of the resistance of \kCl\ located at different positions in the temperature-pressure phase diagram shown in the inset. Due to the stress acting on the sample \kClstar\ embedded in epoxy, it exhibits a semiconducting (insulating) behavior down to about 46\,K below which the resistivity shows a pronounced minimum following the transition into the metallic phase. The S-shaped transition line of the MIT (see inset) is crossed again leading once more to insulating behavior below about 22\,K, where a kink in the resistivity curve is observed. Upon further cooling, the sample undergoes an insulator-to-superconductor transition at $T_c = 13$\,K (midpoint of the transition). The transition is rather broad due to the inhomogeneous nature of the coexistence region. The resistivity drops within $\Delta T_c = 1.9$\,K from 90\,\% to 10\,\% of its normal-state value.
The observed behavior is in good agreement with the phase diagram determined from isobaric temperature and isothermal pressure sweeps by Kagawa {\em et al.}\ \cite{Kagawa2004,Kagawa2005}. The reference samples \kCl\ showed the expected semiconducting behavior down to low temperatures. 

\begin{figure}[]
%\sidecaption
\includegraphics[width=.475\textwidth]{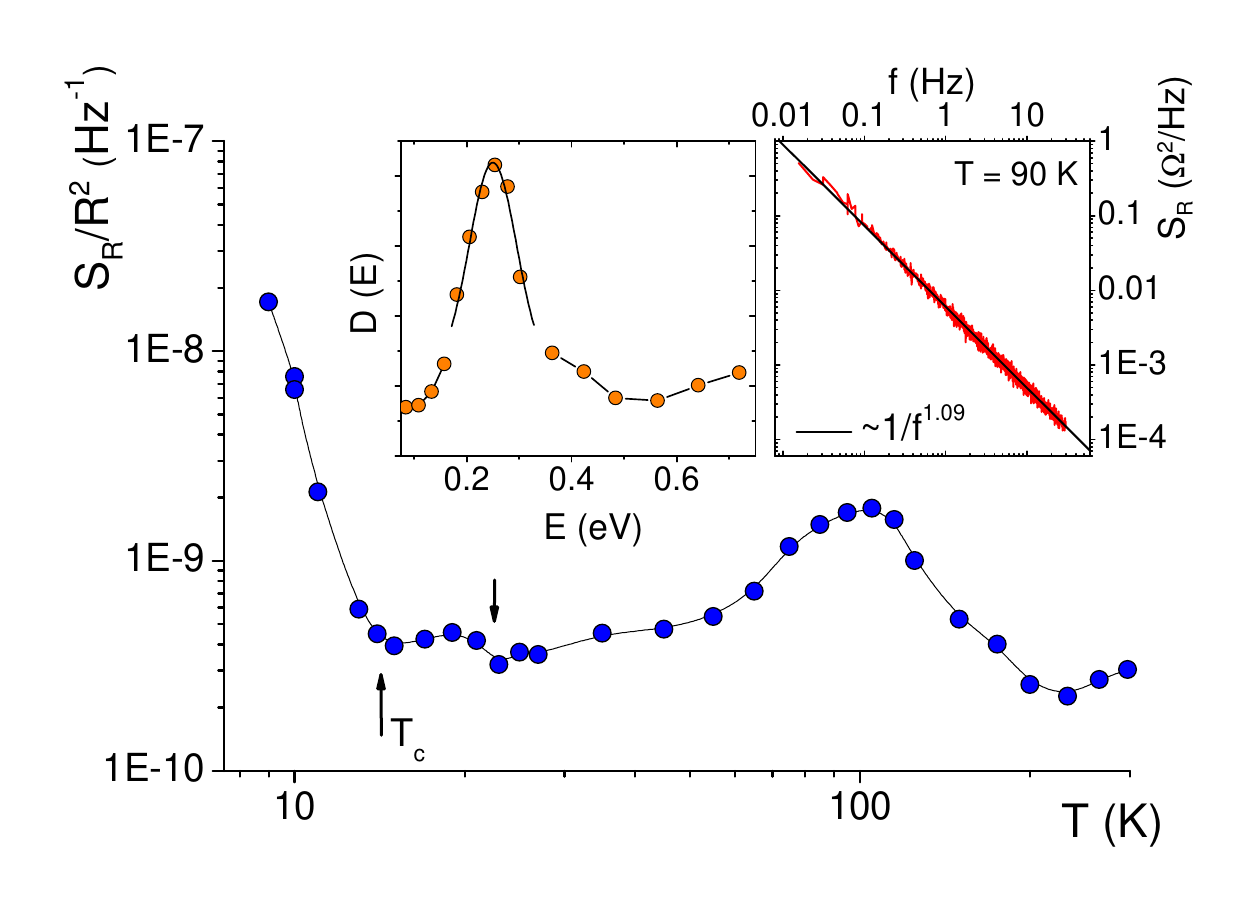}
\caption[]{(Color online.) Temperature dependence of the normalized resistance noise at 1\,Hz of \kClstar. Arrows indicate the superconducting transition (onset of $T_c$) and the temperature of re-entering the insulating phase. Right inset: PSD of the resistance noise at 90\,K in a log-log plot. The line is a linear fit yielding $S_R \propto 1/f^{1.09}$. Left inset: energy distribution $D(E)$ extracted from the DDH model, see text.}\label{figure2}
\end{figure}
A typical noise power spectral density (PSD) of \kClstar\ taken at $T = 90$\,K is shown in Fig.\,2 (right inset). At all temperatures above $T_c$ we observe a noise spectrum of generic $1/f^\alpha$-type with $0.9 < \alpha < 1.1$. The main panel shows the normalized noise $S_R/R^2(T)$ taken at 1\,Hz. We apply a phenomenological random fluctuation model introduced by Dutta, Dimon, and Horn (DDH) \cite{Dutta1979} based on the assumption that $1/f$ spectra are created by the superposition of a large number of random and independent switching entities called "fluctuators", which are assumed to linearly couple to the resistance of the sample:  
\begin{equation}
S(f) \propto \int\frac{\tau(E)}{\tau(E)^2 4 \pi^2 f^2 +1}D(E)dE.
\label{DDHeq1}
\end{equation}
Each fluctuator is characterized by a thermally-activated time constant $\tau = \tau_0 \exp(E/k_B T)$, where $\tau_0$ is a characteristic Òattempt timeÒ of the order of an inverse phonon frequency, $E$ the activation energy of the process and $k_B$ Boltzmann's constant. $D(E)$ represents a certain distribution of activation energies that is deduced from the application of the DDH model to the temperature dependent noise $S_R(T)$ (details of the analysis will be published elsewhere). The resulting $D(E)$ (left inset of Fig.\,\ref{figure2}) exhibits a pronounced, roughly Gauss-shaped peak at an energy of $\sim 250$\,meV. This energy is known as the signature of the orientational degrees of freedom of the ET molecules' terminal ethylene moieties undergoing a glass-like transition at around 70\,K \cite{Mueller2002a,Toyota2007}, which corresponds to a certain degree of disorder. The peak energy agrees well with values of $224 - 232$\,meV reported in the literature \cite{Miyagawa1995,Akutsu2000,Mueller2002a}. This demonstrates that fluctuation spectroscopy provides a promising novel access to the microscopic transitions and dynamical molecular properties of the present materials.\\
Upon further cooling the sample, a small but distinct step-like increase in $S_R/R^2$ is observed at 22\,K (see arrow in Fig.\,\ref{figure2}), which coincides with a kink in the resistivity (see Fig.\,\ref{figure1} and text above). This feature is most likely due to the crossing of the MIT line when re-entering the insulating state, where an increase of the noise due to localization effects is expected. In the following we will focus on the transition into the superconducting state which goes along with a substantial rise of the noise level. We interpret its more than two-orders-of-magnitude increase when cooling through $T_c$ as a result of the percolative nature of the superconducting transition in this sample. Similar behavior has been observed for many high-$T_c$-cuprate samples, see e.g.\ \cite{Testa1988,Lee1989,Kiss1994,Kogan1996} and references therein. Qualitatively, in strongly disordered conductors, the resistance fluctuations are determined not by the entire volume of the conductor but by an essentially smaller volume leading to a large noise level \cite{Weissman1988,Kogan1996}. Alternatively, strongly disordered conductors exhibit much higher resistance noise than homogeneous or weakly disordered ones %with the same characteristics (dimensions and resistance) 
due to the strongly non-uniform current and electric field distribution confined to narrow paths.\\ We now apply ideas of percolation theory and treat the coexistence region close to the MIT as a mixture of superconducting and non-superconducting (normal or insulating) phases, i.e.\ a lattice of resistors with a temperature-dependent fraction $p$ that is short-circuited, simulating the superconducting links. For instance, one can think of a network of $p$ Josephson-coupled junctions formed by connections between superconducting grains or clusters. A wide distribution of junction critical currents $i_c(T)$ (determined by inhomogeneities on the length scale of the superconducting coherence length) means that at a given macroscopic current $I$, the local current $i$ can be either larger or smaller than $i_c$ which determines if the Josephson junction is superconducting or resistive. In this "classical" percolation model \cite{Kiss1993b}, the noise level is determined by fluctuating (non-superconducting) resistors $r_i + \Delta r_i(t)$, i.e.\ resistive junctions. With decreasing temperature, $p$ %--- defined as the fraction of resistors which are short-circuited by superconducting links --- 
increases, so the noisy volume contributing to the macroscopic resistance fluctuations $R +\Delta R(t)$ shrinks, which in turn leads to a divergence of the spectral density of fractional resistance fluctuations when approaching the percolation threshold $p_c$ from above, in agreement with our observations shown in Fig.\,\ref{figure2}.

\begin{figure}[]
%\sidecaption
\includegraphics[width=.475\textwidth]{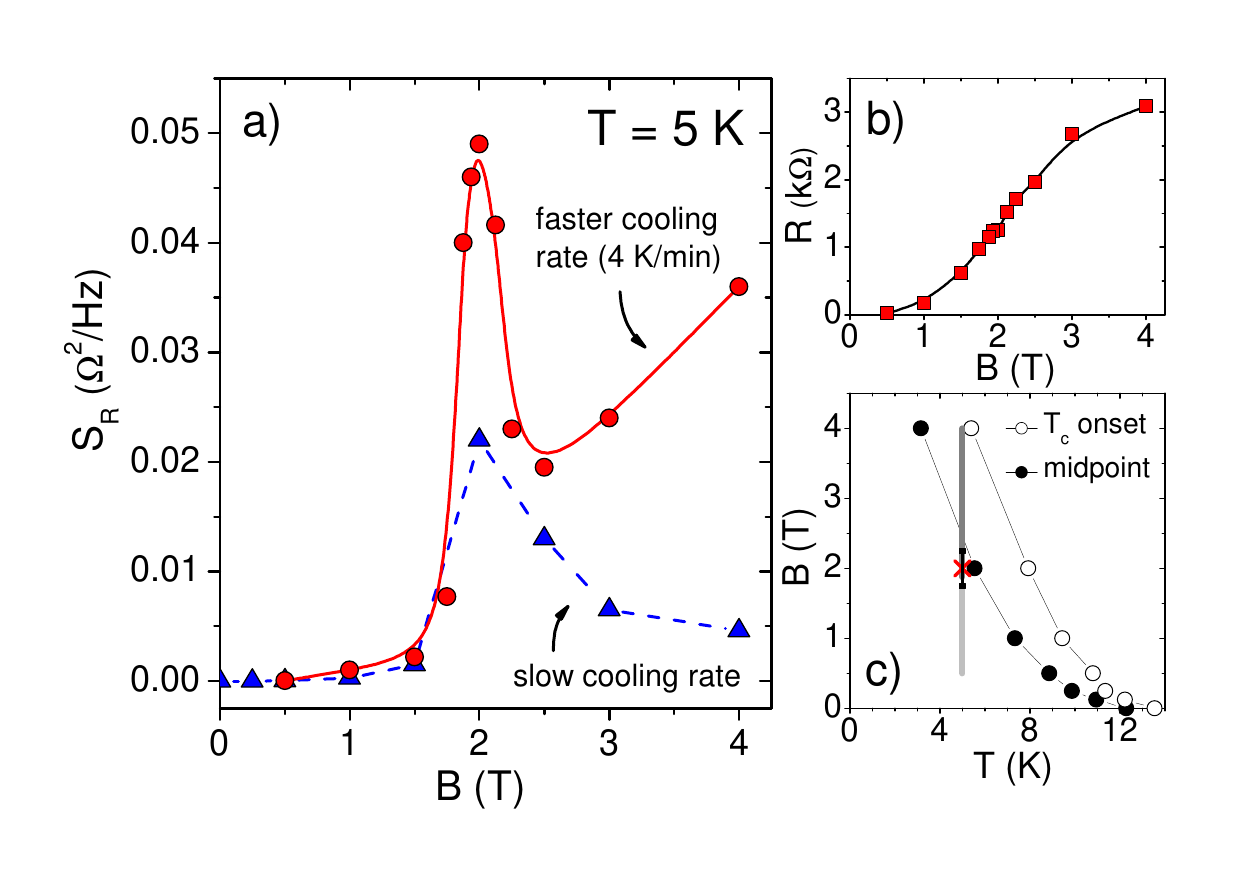}
\caption[]{(Color online.) a) $S_R (B,T = 5\,{\rm K})$ at 1\,Hz for different cooling rates. b) Magnetosresistance at 5\,K and c) phase diagram with the onset and midpoint of the superconducting transition at various fields. Dark and light grey lines indicate different scaling regimes (see text and Fig.\,\ref{figure4}), the cross shows the maximum noise level, and the black line indicates the regime, where the noise deviates from $1/f$-behavior.}\label{figure3}
\end{figure}
In the present materials, even when homogeneous, it is known that the effect of fluctuations is strongly enhanced in increasing magnetic fields \cite{Toyota2007}, which in combination with the percolative nature of superconductivity might enhance the signature of only a few fluctuators. Consequently, we have studied the percolation effects at the superconducting transition in more detail by performing isothermal noise measurements in magnetic fields applied perpendicular to the planes. Fig.\,\ref{figure3} a) shows that the magnitude of the resistance noise at $T = 5$\,K strongly depends on the magnetic field exhibiting a pronounced peak at $B = 2$\,T, just below the midpoint of the superconducting transition, see Fig.\,\ref{figure3} b) and c). Also, the PSD is sensitive to disorder in the ethylene endgroups degrees of freedom which can be tuned in the present compounds by applying different cooling rates \cite{Toyota2007}. Percolation theory predicts power-law scaling behavior in the variable $(p - p_c)$ which very often cannot be determined accurately. Thus, it is more convenient \cite{Kogan1996} to consider a scaling law $S_R/R^2 \propto R^ {l_{rs}}$, where $l_{rs}$ is the resistor-superconductor (RS) network scaling exponent.   
\begin{figure}[]
%\sidecaption
\includegraphics[width=.475\textwidth]{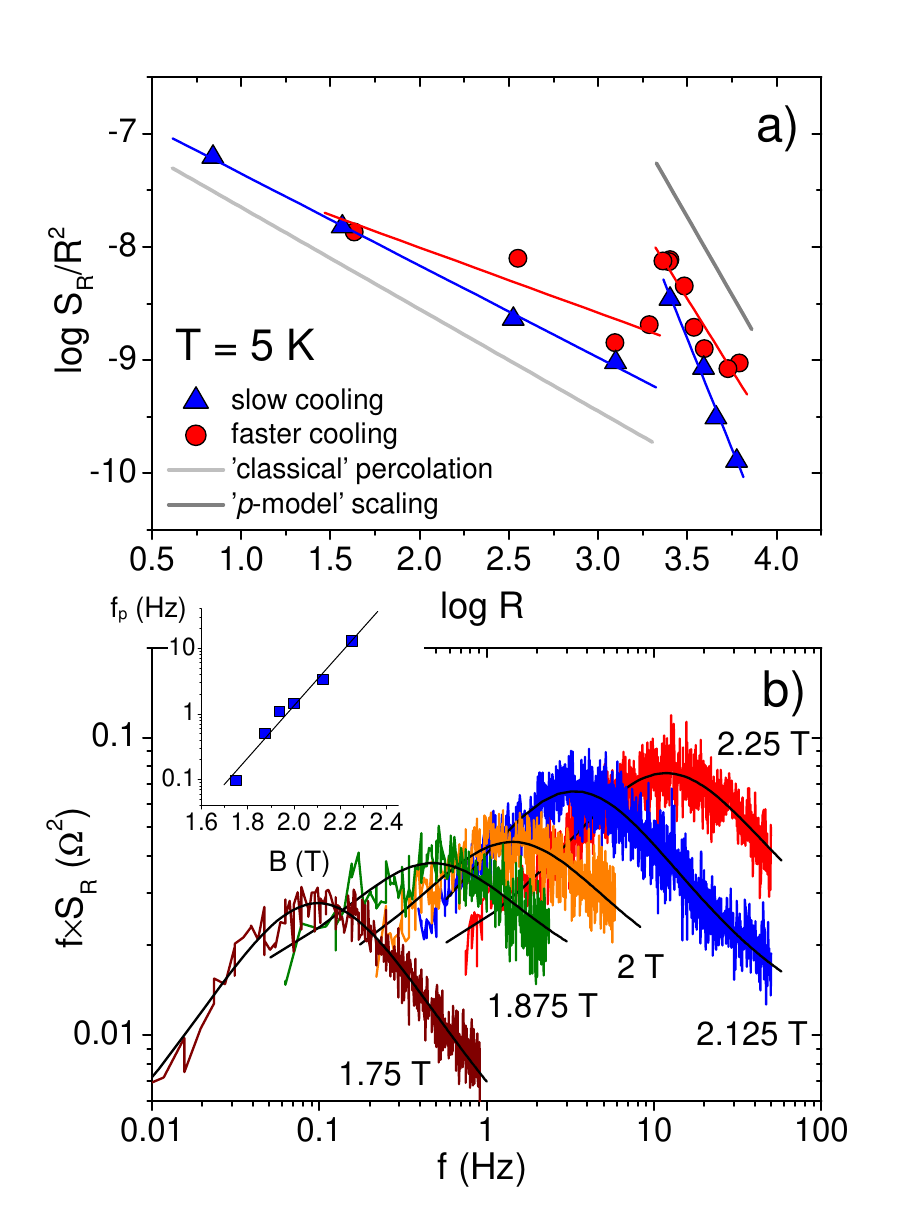}
\caption[]{(Color online.) a) Scaling of the normalized noise $S_R/R^2$ versus the resistance $R$, for the data in Fig.\,\ref{figure3}. b) PSD in the vicinity of the maximum noise level as $f \times S_R$ vs.\ $f$. Lines are fits to eq.\ (\ref{Lorentzian}) plus a $1/f$ background term. Inset shows the shift of the peak frequency $f_p$ with applied magnetic field $B$.}\label{figure4}
\end{figure}
Surprisingly, as shown in Fig.\,\ref{figure4} a), the transition is not just dominated by a single scaling law but rather by two percolation regimes (different slopes $l_{rs}$) with an apparent discontinuity corresponding to the peak in the noise level. Deep in the superconducting phase (low fields), the scaling follows the prediction for a "classical" RS network in 3D of $l_{rs} = 0.9 \pm 0.3$ \cite{Kiss1993b}, whereas the slope when approaching superconductivity from above (high fields) is significantly larger. The slope we observe is close to the characteristic exponent of $l_{rs} = 2.74$ for the so-called $p$-model in 3D \cite{Kiss1993a,Kiss1993b}, which has been first observed for thin-film high-$T_c$ superconductors. $p$-noise, leading to a new class of universal scaling exponents, in the above-described picture corresponds to a random switching (on-off) of superconducting links in the RS network, i.e.\ spontaneous fluctuations $\Delta p(t)$ of $p(T)$ leading to fluctuations in the macroscopic resistance, which may be due to time-dependent perturbations of the Josephson coupling energy between superconducting clusters. When approaching $T_c (B)$ from above, percolation seems to be dominated by unstable superconducting clusters/domains switching back and forth. At a field of 2\,T, which is just below the midpoint of the resistive transition, the noise peaks and shows a crossover to a "classical" percolation regime dominated by the change in the fraction of superconducting domains.
Fig.\,\ref{figure4} b) shows the PSDs as $f \times S_R$ vs.\ $f$ at various fields in the vicinity of 2\,T where the noise peaks. In this narrow field range we observe clear deviations from $1/f$-type behavior which in this representation appears as a constant background. Obviously, under certain conditions ("noise window") the spectra are resolved into Lorentzian contributions superimposed on a $1/f$ background. A Lorentzian PSD 
\begin{equation}
S_R(f) = \frac{4 (\Delta R)^2}{(\tau_1 + \tau_2)[(\tau_1^{-1} + \tau_2^{-1})^2+(2 \pi f)^2]}
\label{Lorentzian}
\end{equation}
represents a random telegraph signal between "high" and "low" resistance levels in the time domain  \cite{Machlup1954}, where $\tau_i$ denote the characteristic lifetimes for the "high" and "low" states. Lines in Fig.\,\ref{figure4} b) are fits to a $1/f$ background plus eq.\ (\ref{Lorentzian}) showing that an individual fluctuator dominates the noise. For the relative resistance change we infer $\Delta R/R \sim 0.2$\,\%. The inset of Fig.\,\ref{figure4} shows a clear shift of the peak frequency $f_p = 1/2\pi (\tau_1^{-1} + \tau_2^{-1})$ to higher values with increasing magnetic field. One may consider the fluctuating entity as a cluster that switches between the superconducting and normal state and assume a thermally activated behavior, for which at a given temperature the energy barrier depends on the magnetic field: $f_p = \tau_0^{-1} \exp[-(E + mB)/k_B T]$. Here, $E$ denotes the energy barrier at zero field and $mB$ the magnetic energy, for which we find a value of $\sim 3.45$\,meV from the fit to the data, see inset of Fig.\,\ref{figure4}. If we compare this energy to the condensation energy density $V_s B^2_{c_{th}}/2 \mu_0$, where $B_{c_{th}}$ is the thermodynamic critical field and $\mu_0$ the free space permeability, we get a rough estimate for the fluctuating sample volume of $V_s \sim 340\,{\rm nm}^3$ \cite{comment}. The estimated volume indicates a mesoscopic rather than a macroscopic inhomogeneous state. An alternative explanation would be that in an already superconducting cluster the vortices become temporarily pinned and unpinned, possibly going along with switching between vortex solid and liquid state. Assuming that in such a case $mB$ may simply be compared to the pinning energy %$n \epsilon^\ast V_s$, where $n$ is the number of flux lines per unit area, $\epsilon^\ast = B_{c_1} \Phi_0 / \mu_0$ the energy per unit length of vortex ($\Phi_0$ is the flux quantum), 
we find a value for $V_s$ which is smaller but in the same order as the one above.\\ 
In summary, the low-frequency electronic fluctuations in the spatially inhomogeneous coexistence state in pressurized $\kappa$-(ET)$_2$Cu[N(CN)$_2$]Cl indicate a crossover between different percolation regimes for the superconducting transition in finite magnetic fields. The data can be explained by a random-resistor network
consisting of superconductung and resistive domains for which we estimate a cluster volume in the order of a few hundred nm$^3$. 

\begin{acknowledgments}
Work supported by the Deutsche Forschungsgemeinschaft (DFG) through the Emmy Noether program. J.M. is grateful to Dr.\ S.\ Wirth for helpful discussions. 
Work at Argonne National Laboratory is sponsored by the U. S. Department of Energy, Office of Basic Energy Sciences, Division of Materials Sciences, under Contract DE-AC02-06CH11357.
\end{acknowledgments}

%\bibliography{noise1}% Produces the bibliography via BibTeX.

\end{document}